\PassOptionsToPackage{svgnames}{xcolor}

\documentclass[sigconf,screen]{acmart}

\AtBeginDocument{%
  }

\setcopyright{none}
\renewcommand\footnotetextcopyrightpermission[1]{}

\graphicspath{{figs/}{figures/}{pictures/}{images/}{./}} 

\usepackage{booktabs}                  
\usepackage[normalem]{ulem}

\usepackage{url}
\usepackage{tabularx}

\definecolor{RevisionPink}{HTML}{D81B60}

\newcommand{\rev}[1]{{\color{black}#1}}
\newenvironment{revisionblock}{\begingroup\color{black}}
{\endgroup}
\usepackage[most]{tcolorbox}
\usepackage{enumitem}

\definecolor{BG}{HTML}{ffffff}
 \newtcbox{\inlineMainCategoryBox}[2][]{enhanced,
 box align=base,
 nobeforeafter,
 colback=#2!45!BG, 
 colframe=#2!100!BG, 
 size=small,
 toprule=0pt,
 bottomrule=0pt,
 leftrule=0pt,
 rightrule=0pt,
 fontupper=\footnotesize,
 left=1pt,
 right=1pt,
 boxsep=1.2pt,
 #1}

\newtcbox{\inlineSubCategoryBox}[2][]{enhanced,
 box align=base,
 nobeforeafter,
 colback=#2!30!BG,
 colframe=#2!30!BG,
 size=small,
 toprule=0pt,
 bottomrule=0pt,
 leftrule=0pt,
 rightrule=0pt,
 fontupper=\footnotesize,
 left=1pt,
 right=1pt,
 boxsep=1.2pt,
 #1}

\newcommand{\executionTitle}[0]{{\color{Execution} \textbf{Execution}}}
\newcommand{\designTitle}[0]{{\color{Design} \textbf{Design}}}
\newcommand{\responseTitle}[0]{{\color{Response} \textbf{Response}}}
\newcommand{\objectiveTitle}[0]{{\color{Objective} \textbf{Objective}}}
\newcommand{\promptTitle}[0]{{\color{Prompt} \textbf{Traits}}}

\newcommand{\executionHeading}[0]{\textcolor{Execution}{Execution}}
\newcommand{\designHeading}[0]{\textcolor{Design}{Design}}
\newcommand{\responseHeading}[0]{\textcolor{Response}{Response}}
\newcommand{\objectiveHeading}[0]{\textcolor{Objective}{Objective}}
\newcommand{\promptHeading}[0]{\textcolor{Prompt}{Prompt Traits}}

\newcommand{\executionSub}[1]{{\inlineMainCategoryBox{Execution}{\textbf{#1}}}}
\newcommand{\executionCode}[1]{{\inlineSubCategoryBox{Execution}{\textit{#1}}}}

\newcommand{\designSub}[1]{{\inlineMainCategoryBox{Design}{\textbf{#1}}}}
\newcommand{\designCode}[1]{{\inlineSubCategoryBox{Design}{\textit{#1}}}}

\newcommand{\responseSub}[1]{{\inlineMainCategoryBox{Response}{\textbf{#1}}}}
\newcommand{\responseCode}[1]{{\inlineSubCategoryBox{Response}{\textit{#1}}}}

\newcommand{\objectiveSub}[1]{{\inlineMainCategoryBox{Objective}{\textbf{#1}}}}
\newcommand{\objectiveCode}[1]{{\inlineSubCategoryBox{Objective}{\textit{#1}}}}

\newcommand{\promptSub}[1]{{\inlineMainCategoryBox{Prompt}{\textbf{#1}}}}
\newcommand{\promptCode}[1]{{\inlineSubCategoryBox{Prompt}{\textit{#1}}}}

\definecolor{Execution}{HTML}{469a93}
\definecolor{Design}{HTML}{2c7a32}
\definecolor{Response}{HTML}{1e5ca9}
\definecolor{Objective}{HTML}{DA6B46} 
\definecolor{Prompt}{HTML}{e0a503}

\newcommand{\newcategory}[2]{%
  \definecolor{#1_color}{HTML}{#2}%
  \expandafter\newcommand\csname #1\endcsname[1]{%
    {\setlength{\fboxsep}{1.5pt}
    \colorbox{#1_color}{\textcolor{black}{\texttt{##1}}}}%
  }%
}
\newcommand{\newsubcategory}[2]{%
  \definecolor{#1_color}{HTML}{#2}%
  \expandafter\newcommand\csname #1\endcsname[1]{%
    {\setlength{\fboxsep}{1pt}
    \colorbox{#1_color}{\textcolor{black}{\texttt{##1}}}}%
  }%
}
\newcategory{execution}{A0C5C2}
\newsubcategory{executionb}{CFE1E0}
\newcategory{chartdesign}{9ECAA3}
\newsubcategory{chartdesignb}{CDE4D0}
\newcategory{response}{9DBFD0}
\newsubcategory{responseb}{CDDEE7}
\newcategory{initiation}{ECB3A2}
\newsubcategory{initiationb}{F6D8CF}
\newcategory{objective}{F7BF96}
\newsubcategory{objectiveb}{FBDFC9}
\newcategory{prompttraits}{F0D18A}
\newsubcategory{prompttraitsb}{F8E8C3}

\makeatother

\begin{document}

\title{Vibe Visualizing: How Visualization Novices Try (and Fail) to Generate and Interpret Visualizations with Conversational AI}

\author{Sam Yu-Te Lee}
\orcid{0000-0002-1825-0097}
\affiliation{%
  \institution{University of California, Davis}
  \city{Davis}
  \state{California}
  \country{USA}}
\email{ytlee@ucdavis.edu}

\author{Yun-Hsin Kuo}
\affiliation{%
  \institution{University of California, Davis}
  \city{Davis}
  \state{California}
  \country{USA}}
\email{yskuo@ucdavis.edu}

\author{Chifang Chou}
\affiliation{%
  \institution{University of California, Davis}
  \city{Davis}
  \state{California}
  \country{USA}}
\email{cfchou@ucdavis.edu}

\author{Matthew Ward}
\affiliation{%
  \institution{University of California, Davis}
  \city{Davis}
  \state{California}
  \country{USA}}
\email{msward@ucdavis.edu}

\author{Xiwei Xuan}
\affiliation{%
  \institution{University of California, Davis}
  \city{Davis}
  \state{California}
  \country{USA}}
\email{xwxuan@ucdavis.edu}

\author{Kwan-Liu Ma}
\affiliation{%
  \institution{University of California, Davis}
  \city{Davis}
  \state{California}
  \country{USA}}
\email{klma@ucdavis.edu}

\renewcommand{\shortauthors}{Lee et al.}

\begin{abstract}
\rev{
Conversational AI lowers technical barriers to generating and interpreting visualizations. People without visualization training or practical experience can therefore perform these tasks, but may lack the expertise needed to evaluate the resulting charts and interpretations. Through a study with 20 visualization novices, we examine conversation logs from 60 sessions with ChatGPT and develop a codebook that characterizes participant and AI behaviors in AI-assisted visualization. Paired with interview transcripts and survey responses, we derive three themes under outcome quality, AI non-compliance, and trust, verification, and repair. An exploratory role analysis identifies two session-level roles distinguished by user initiative and associated with different failure patterns. A comparative analysis with Gemini and Claude further reveals distinct system-specific behaviors. Based on these findings, we derive five design recommendations for AI-assisted visualization design and discuss implications for visualization literacy, human--AI roles, accountability, and agentic visualization. 
}
\end{abstract}




\maketitle

\pagestyle{plain}
\thispagestyle{plain}

\section{Introduction}
\rev{
Conversational AI enables people without formal training or practical experience in visualization to generate and interpret visualizations through natural language. We call this practice \emph{vibe visualizing}, by analogy to vibe coding~\cite{karpathy2025vibecoding}, and refer to these users as visualization novices. 
General-purpose conversational AI systems such as ChatGPT~\cite{chatgpt} now make this practice widely accessible.
However, current systems can generate flawed or misleading visualizations~\cite{tian2025chartgpt} and misinterpret visualizations~\cite{hong2025LLM_VL_modified_VLAT,dong2025VL_VLM}. 
Such failures can be difficult to verify even for experienced data analysts~\cite{Gu2024data_analyst_verify}, raising particular concern for novices who are less prepared to evaluate analysis and design decisions determined by AI.
}
%

\rev{
Visualization generation and interpretation are core components of conversational visual analytics, but research have often examined them separately
~\cite{gu2026i_need_chart_data}.}
Recent work has demonstrated the feasibility of generating visualizations through code generation~\cite{sah2024nl4dvllm, tian2025chartgpt}, built in Large Language Models (LLMs), yet methods for evaluating and improving these visualizations remain underexplored~\cite{basole2024genai_vis}. Consequently, it remains unclear how reliably LLMs can generate visualizations that well support user goals.
In contrast, research on visualization interpretation, facilitated by Vision Language Models (VLMs), has established benchmarks for systematically evaluating visualization literacy in AI~\cite{mukherjee2025encqa, liu2025simvecvis, masry2025chartqapro}.
However, these systems continue to struggle across analysis tasks, chart types, and prompting methods~\cite{dong2025VL_VLM, hong2025LLM_VL_modified_VLAT}. 
\rev{
Existing benchmarks also do not fully capture the ambiguity of real-world interpretations~\cite{knoll2025gulf_of_interpretation,bearfield2024same_data_diverging_perspectives} or the challenges posed by deceptive and poorly designed visualizations~\cite{Mahbub2025misleading_vis_VLM}.
}
\rev{
These limitations raise an important question: what happens when visualization novices need to leverage visualizations for various purposes and resort to visualizations and interpretations produced by a conversational AI system? We are therefore motivated to investigate what problems arise in these interactions, whether visualization novices recognize them, and how they respond.
}
\rev{In this work, we conduct a user study in which 20 visualization novices complete 60 sessions with ChatGPT.}
\rev{We analyze conversation logs, interview transcripts, documented insights, and post-task ratings. Three coders experienced in visualization design develop a 64-code codebook covering AI execution, chart design issues, AI response behaviors, and participant prompting behaviors. Our thematic analysis identifies three themes: failures in chart and insight quality; AI non-compliance and recurring specification difficulties; and trust, limited verification, and failed repair loops. Through an exploratory behavioral analysis of human--AI interaction in our study, we identify two session-level roles characterized by different levels of user initiative. In Delegator sessions, participants leave analytical and design decisions to ChatGPT; in Decision-maker sessions, participants specify or question those decisions.}

\rev{To examine whether our codebook also captures similar behaviors and failures in other convesational AI systems, we replay the 60 opening prompts with Gemini~\cite{gemini} and Claude~\cite{claude} and annotate their responses following the same procedure. Our comparative analysis reveals distinct system-specific response and failure profiles.}
Based on the findings across these analyses, we derive five design recommendations for AI-assisted visualization systems. \rev{We conclude by discussing implications for interdependent literacy gaps, shifting human--AI roles, accountability, and agentic visualization.}
\rev{We consider our contributions to be:}
\begin{itemize}[noitemsep]
    \item \rev{Empirical evidence of the problems visualization novices encounter when generating and interpreting visualizations with conversational AI, including flawed outputs, limited verification, and failed repair attempts;}
    \item \rev{A 64-code codebook characterizing novice and AI behaviors in AI-assisted visualization, together with a thematic analysis that identifies recurring patterns;}
    \item \rev{A broader account of how interaction failures vary and can be addressed through (1) an exploratory analysis of session-level roles, (2) a comparison across ChatGPT, Gemini, and Claude, and (3) five system design recommendations.}

\end{itemize}





\section{Related Work}
\rev{Our work connects research on natural-language interfaces for visualization, visualization interpretation and literacy, and appropriate reliance on conversational AI.}

\subsection{\texorpdfstring{\rev{Natural-Language Interfaces for Visualization}}{Natural-Language Interfaces for Visualization}}
\begin{revisionblock}
Visualizations are conventionally created through code, such as Matplotlib~\cite{Hunter2007matplotlib} and D3.js~\cite{bostock2011d3}; expressive specifications, such as Vega-Lite~\cite{satyanarayan2017vega_lite}; or specialized tools, such as Tableau~\cite{tableau2026} and Power BI~\cite{powerbi2026}. Natural-language interfaces lowered part of this technical barrier before the emergence of LLMs. Eviza supported iterative questions about an existing visualization and exposed widgets for resolving ambiguous terms~\cite{setlur2016eviza}. NL4DV translated natural-language queries into analytic specifications containing data attributes, tasks, and recommended visualizations~\cite{narechania2021nl4dv}. Tableau's Ask Data similarly returned interactive visualizations in response to natural-language questions over a prepared data source~\cite{tableau2018askdata}.

LLMs extend this interaction model by generating visualizations from less constrained requests~\cite{sah2024nl4dvllm,tian2025chartgpt} and supporting iterative refinement through conversation~\cite{dhanoa2025agentic_visualization,zhao2025light_va,chen2025interchat}. Public-facing systems such as ChatGPT~\cite{chatgpt}, Gemini~\cite{gemini}, and Claude~\cite{claude} combine visualization generation, analysis, and interpretation in a single interface. This form of end-to-end delegation parallels vibe coding~\cite{karpathy2025vibecoding}: users can express an intended outcome while leaving implementation details to the system. However, LLM-based visualization systems can fail to follow instructions~\cite{zhao2025light_va}, provide insufficient design justification~\cite{dibia2023lida}, and make inconsistent design decisions~\cite{maddigan2023Chat2Vis}. Prior systems and technical evaluations establish the feasibility of natural-language visualization, but leave open how novices evaluate the analysis and encoding decisions that a general-purpose conversational AI makes on their behalf.
\end{revisionblock}

\subsection{\texorpdfstring{\rev{Visualization Interpretation, Design Feedback, and Literacy}}{Visualization Interpretation, Design Feedback, and Literacy}}
\begin{revisionblock}
Recent work has evaluated VLMs on chart question answering and standardized visualization tasks~\cite{dong2025VL_VLM,hong2025LLM_VL_modified_VLAT,das2025charts_of_thought}. Performance varies substantially across chart types, analysis tasks, models, and prompting methods~\cite{hong2025LLM_VL_modified_VLAT}. Documented errors include inaccurate value extraction, incorrect identification of visual patterns, and unsupported conclusions drawn from correctly identified chart elements~\cite{dong2025VL_VLM}. VLMs may also rely on knowledge from their training data instead of visible evidence~\cite{hong2025LLM_VL_modified_VLAT}, and deceptive elements such as truncated or inverted axes and unjustified 3D effects further reduce their reliability~\cite{Mahbub2025misleading_vis_VLM}. At the same time, VLM-generated visualization critiques can provide broad, task-oriented design feedback that experienced practitioners find useful~\cite{Ahn2025ChatGPT_design_advice,shin2025visualizationary}.

Human visualization literacy is also multidimensional. VLAT measures comprehension and interpretation of well-formed visualizations~\cite{lee2017vlat}, and Mini-VLAT provides a shorter assessment of similar abilities~\cite{pandey2023minivlat}. CALVI focuses on critically reading and reasoning about erroneous or potentially misleading visualizations~\cite{ge2023calvi}, while AVEC measures visual encoding ability as one component of visualization construction~\cite{ge2025avec}. These instruments operationalize important parts of visualization reading, critical interpretation, and construction, but visualization-literacy measurement remains context-sensitive and subject to validity threats~\cite{ge2026wicked}. None of these instruments directly assesses the AI-assisted visualization tasks studied here, which require people to communicate a visualization goal with AI, evaluate the AI's data analysis and design choices, interpret the resulting chart, and request revisions.
\end{revisionblock}

\subsection{\texorpdfstring{\rev{Trust, Appropriate Reliance, and Conversational AI}}{Trust, Appropriate Reliance, and Conversational AI}}
\begin{revisionblock}
Trust shapes whether people rely on automation, particularly when they cannot fully understand a complex system. Appropriate reliance therefore requires trust to correspond to the system's capabilities and the context in which it operates~\cite{lee2004trust}. Conversational AI complicates this relationship. Non-AI experts use inconsistent prompts, form mismatched expectations of system capabilities~\cite{zamfirescu2023whyjohnny}, and do not consistently evaluate model responses~\cite{kim2024evallm}. Even experienced data analysts report difficulty understanding an AI's reasoning and verifying its outputs~\cite{Gu2024data_analyst_verify,gu2026i_need_chart_data}.

Prior work has investigated ways to support more appropriate reliance, including supplying sources and supporting details~\cite{Kim2025explanations_sources_inconsistencies}, expressing uncertainty~\cite{Kim2024uncertainty_expression}, communicating confidence~\cite{radensky2023confidence_signal_patterns}, and using cognitive forcing functions that require people to reason before seeing an AI recommendation~\cite{buccinca2021cognitiveforcing}. These interventions can encourage scrutiny, but they generally presume that users can evaluate the evidence or formulate an independent judgment. That presumption is especially consequential in vibe visualizing: the generated chart is both an AI output that must be evaluated and the evidence through which the AI's interpretation is supposed to be checked. Existing research therefore leaves open how trust and reliance develop when visualization novices must evaluate the artifact, the claims derived from it, and the repair attempts produced by the same conversational system.
\end{revisionblock}

\begin{table*}[t]
\centering
\caption{Scenario-based tasks in our user study. Detailed task instructions, participant conversation logs, and coding results can be found at \url{https://vibe-visualizing.vercel.app/}}
\renewcommand{\arraystretch}{1.2}
\setlength{\extrarowheight}{0.5pt}
\begin{tabularx}{\textwidth}{l | >{\raggedright\arraybackslash}p{3.3cm} >{\raggedright\arraybackslash}X}
\toprule
\textbf{Dataset} & \textbf{Expected Data Patterns} & \textbf{The Selected Most Effective Design}\\
\midrule
Movies
& Examine relationships between budget and gross across genres
& \rev{Scatterplots of budget versus gross, faceted by genre}\\
\cline{2-3}
& Compare return on investment (ROI) across genres
& \rev{Bar chart of mean or median ROI by genre on a logarithmic scale}\\
\cline{2-3}
& Describe long-term trends in budget and revenue
& Line chart of mean or median budget and revenue over time\\
\midrule
Cars
& \rev{Examine relationships among horsepower, miles per gallon (MPG), and weight}
& \rev{Scatterplot of horsepower versus MPG, colored by weight}\\
\cline{2-3}
& Describe changes in MPG over time across regions
& Line chart of MPG over time, grouped by region\\
\cline{2-3}
& \rev{Compare two metrics across geographic regions}
& Dual-axis bar chart, bar-line chart, or box plot\\
\bottomrule
\end{tabularx}
\label{tab:tasks}
\Description{Table with two dataset groups, Movies and Cars, each containing three tasks. Each row pairs an expected analytical pattern with a selected effective visualization. Relationships use scatterplots; temporal trends use line charts; and group comparisons use bar charts, a bar--line chart, or a box plot. ROI means return on investment, and MPG means miles per gallon.}
\end{table*}
\section{User Study Design}
\rev{We conducted a remote user study to investigate how visualization novices used conversational AI to generate and interpret visualizations. To identify issues in this interaction process, we analyzed participants' conversation logs and the visualizations generated during the sessions. We also conducted semi-structured interviews to understand participants' experiences and strategies when interacting with conversational AI. This mixed analysis informed our design recommendations for more robust and user-friendly conversational visualization systems in \autoref{sec:guidelines}.}

\paragraph{Participants} \rev{We operationalized \textbf{visualization novices} using an experience-based recruitment criterion: the absence of formal training or professional experience in data science, data visualization, or graphic design, and of experience with visualization programming (R or Python) or authoring tools (Excel, Tableau, or Power BI).}
\rev{Using this criterion, we recruited 25 individuals through a university's online channels, including Reddit and email lists. We excluded 5 because of technical issues or insufficient English proficiency, resulting in a final sample of 20 participants (13 self-described as men and 7 as women; aged 18--30).}
\rev{The sample included 14 students and 6 working professionals with backgrounds spanning engineering and the social sciences.}
\rev{16 participants reported prior experience using conversational AI, primarily ChatGPT, for everyday tasks such as writing assistance.}

\begin{figure*}[t]
    \centering
    \includegraphics[width=\textwidth]{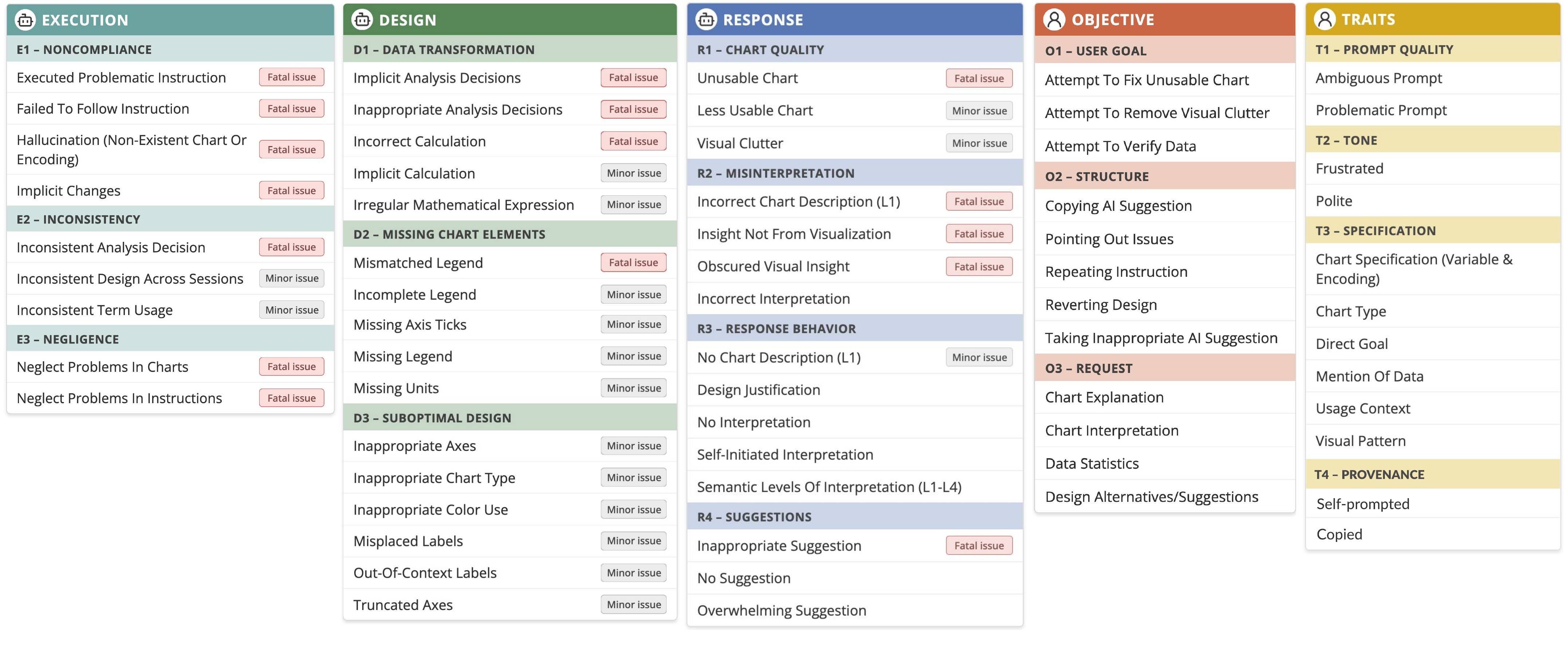}
    \caption{Our codebook, consisting of 64 codes across \rev{5} categories and \rev{17} subcategories, \rev{was developed} through analysis of 60 ChatGPT sessions with 20 visualization novices. \executionTitle{}, \designTitle{}, and \responseTitle{} capture AI-related behaviors in ChatGPT responses, while \objectiveTitle{} and \promptTitle{} capture user-related behaviors in participant prompts. Where applicable, AI-related codes are further labeled as either fatal issues, which can lead to invalid visualizations or interpretations, or minor issues, which reduce the quality of otherwise valid visualizations.}
    \label{fig:codebook}
    \Description{Hierarchical codebook arranged in five columns. Execution, Design, and Response list codes for artificial-intelligence behaviors and outputs; Objective and Traits list codes for participant behaviors and prompts. Each column is divided into labeled subcategories. Badges beside applicable artificial-intelligence codes identify fatal issues and minor issues.}
\end{figure*}
\paragraph{Task and Survey Design}
\rev{
We designed the study around a realistic setting for visualization-driven data analysis.
}
\rev{We chose ChatGPT~\cite{chatgpt} because of its broad public adoption and our participants' familiarity with it.}
\rev{
To collect meaningful interaction data, we selected two familiar datasets (movies and cars) that prior work had shown to be tractable for AI-assisted visualization systems~\cite{sah2024nl4dvllm, tian2025chartgpt, srinivasan2021nlvcorpus}, allowing us to focus on novice–AI interaction rather than failures caused solely by complex data. Each dataset was randomly assigned to 10 participants.
}

\rev{For each dataset, we designed 3 tasks situated in real-world usage scenarios}, e.g., ``\textit{You are building a dashboard for a film investor comparing genres. Use the data to show which genres typically perform better.}''
The tasks implicitly require participants to explore common data patterns, such as trends, comparisons, and relationships between variables, as shown in~\autoref{tab:tasks}.
\rev{The tasks were designed to encourage natural interactions with ChatGPT in which analytical goals were expressed in non-technical language.}
\begin{revisionblock}
As the effectiveness of a visualization design depends on the analysis task and associated context~\cite{saket2019task}, for each task, each of the co-authors independently explored different designs by iterating through data attributes, data processing, and visual encodings.
They identified the most effective design for each task by discussing how well each design presents crucial data patterns, until agreement was reached. 
The resulting set confirmed that each task could be accomplished with standard statistical charts; these designs were not shown to participants and were not treated as the only acceptable design. 
When evaluating participant-generated charts, we considered a design successful so long as it addressed the task and followed established visualization design guidelines~\cite{saket2019task, ge2023calvi, Lan2025design_flaws}.
\end{revisionblock}

\begin{revisionblock}
Besides interacting with ChatGPT, participants completed a post-task survey after each task and a post-study survey, all using 5-point Likert scales. The items in both surveys were informed by the human--AI interaction guidelines by Amershi et al.~\cite{Amershi2019Guidelines} and focused on perceived accuracy, verification need, confidence, and satisfaction. The exact questions and responses are provided in the supplementary materials.
\end{revisionblock}

\rev{The study design and procedure were finalized after 2 pilot tests. First, 1 co-author reviewed the designs, including scenario prompts and survey questions. Then, a volunteer novice participant completed the study to confirm that the procedure was smooth and clear.}


\paragraph{Procedure}
\begin{revisionblock}
We conducted all sessions remotely over Zoom. Participants used their own computers and browsers with access to ChatGPT Plus provided by the research team. The study began with an introduction to the research background and a brief demonstration of ChatGPT's file-upload and visualization features using a third dataset aside from the two datasets used in this study. During the session, we provided the account credentials, the assigned dataset file, and an instruction document containing the three task descriptions.

We instructed participants to start a new chat for each task and complete the 3 tasks in a fixed order. We acknowledge that this design may have introduced ordering effects. However, our analyses characterize recurring patterns across the full corpus rather than compare tasks, and the principal failure patterns appeared across all 3 task positions.
\end{revisionblock}
For each task, participants freely interacted with ChatGPT under no time constraints.
Once they were satisfied with the results, participants proceeded to document 3--5 insights, either by interpreting the generated charts themselves or by using insights provided by ChatGPT, and then completed the post-task survey.
\rev{After all three tasks were completed, participants completed the post-study survey and shared their overall impressions of ChatGPT in a semi-structured interview.}
\rev{Each study session lasted approximately an hour.}
\rev{We collected all ChatGPT conversation logs, screen and audio recordings, and survey responses for further analysis.}
\rev{All participants received a \$20 Amazon gift card within 1 week of participation.}
Our study was determined to be exempt from review by our Institutional Review Board (IRB), and we obtained informed consent from all participants prior to the study.

\section{Coding and Theme Construction}
\label{sec:coding}
\rev{We conducted a thematic analysis of two primary data sources: (1) participants' ChatGPT conversation logs including the generated chart images, and (2) audio transcripts from the entire study.}
\rev{
We analyzed these sources together for their complementary insights; the former records participant actions and AI responses, while the latter captures what participants intend to achieve and how they react to AI responses.
}
All data were anonymized for analysis.

\begin{figure*}
    \centering
\includegraphics[width=\textwidth]{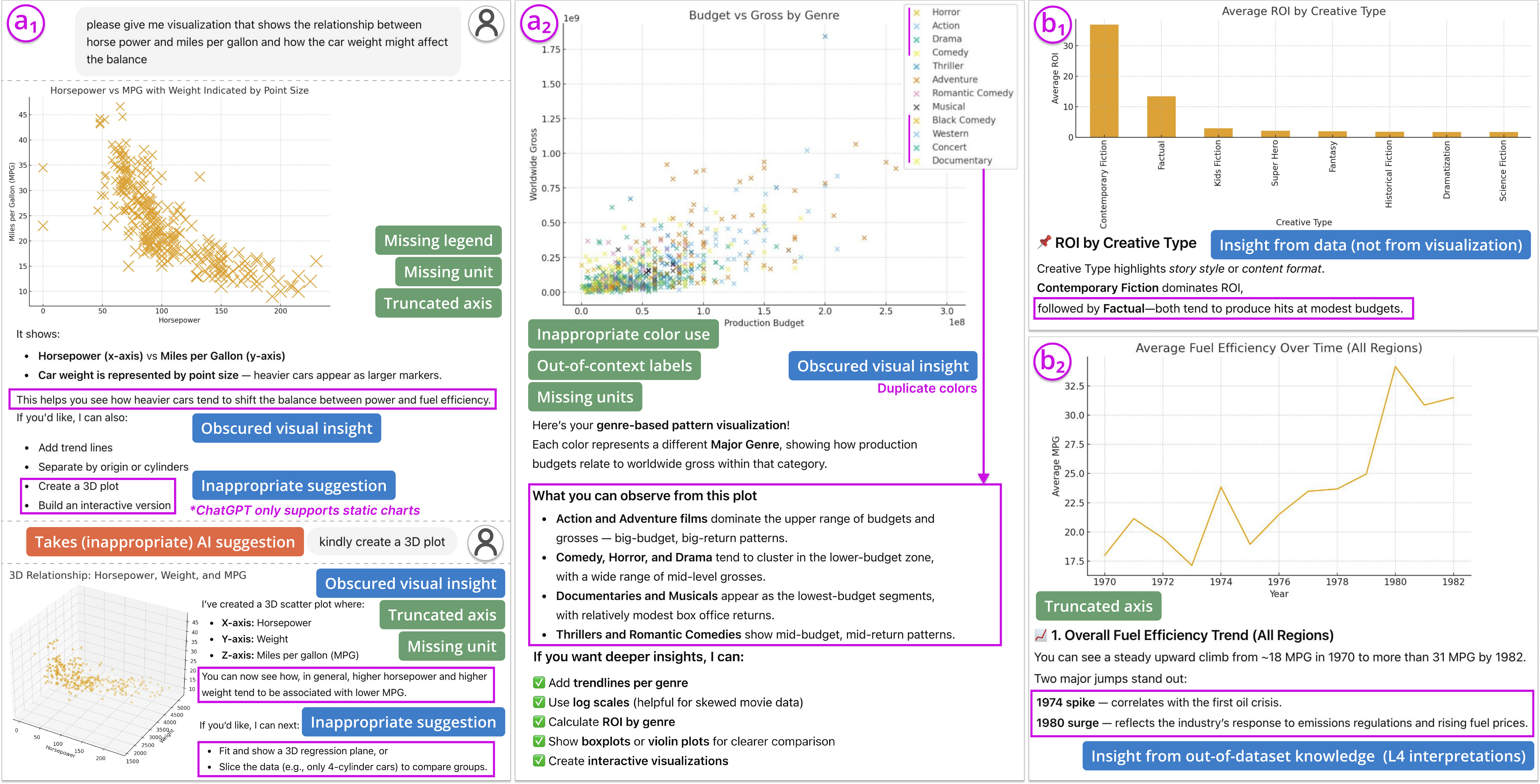}
    \caption{\rev{Examples of ChatGPT-generated charts annotated with codes from the \designTitle{}, \responseTitle{}, and \objectiveTitle{} categories. In (a1), the participant attempts to address visual clutter by following a ChatGPT suggestion; in (a1--a2), duplicate colors and visual clutter make ChatGPT's insights difficult to verify. In (b1--b2), ChatGPT provides insights that are not grounded in the visualizations. Images are resized and rearranged for presentation.}}
    \label{fig:examples_1}
    \Description{Three annotated ChatGPT examples arranged from left to right. Panel a1 shows a two-turn exchange: a cluttered horsepower-versus-fuel-efficiency scatterplot is followed by a more difficult-to-read three-dimensional plot after the participant accepts a suggestion. Panel a2 shows a dense budget-versus-gross scatterplot in which repeated genre colors make the accompanying claims difficult to trace to marks. Panels b1 and b2 pair simple bar and line charts with claims that rely on information not visible in those charts. Callouts identify coded issues, and outlines connect them to the relevant chart or response region.}
\end{figure*}

\begin{figure*}[t]
    \centering
\includegraphics[width=\textwidth]{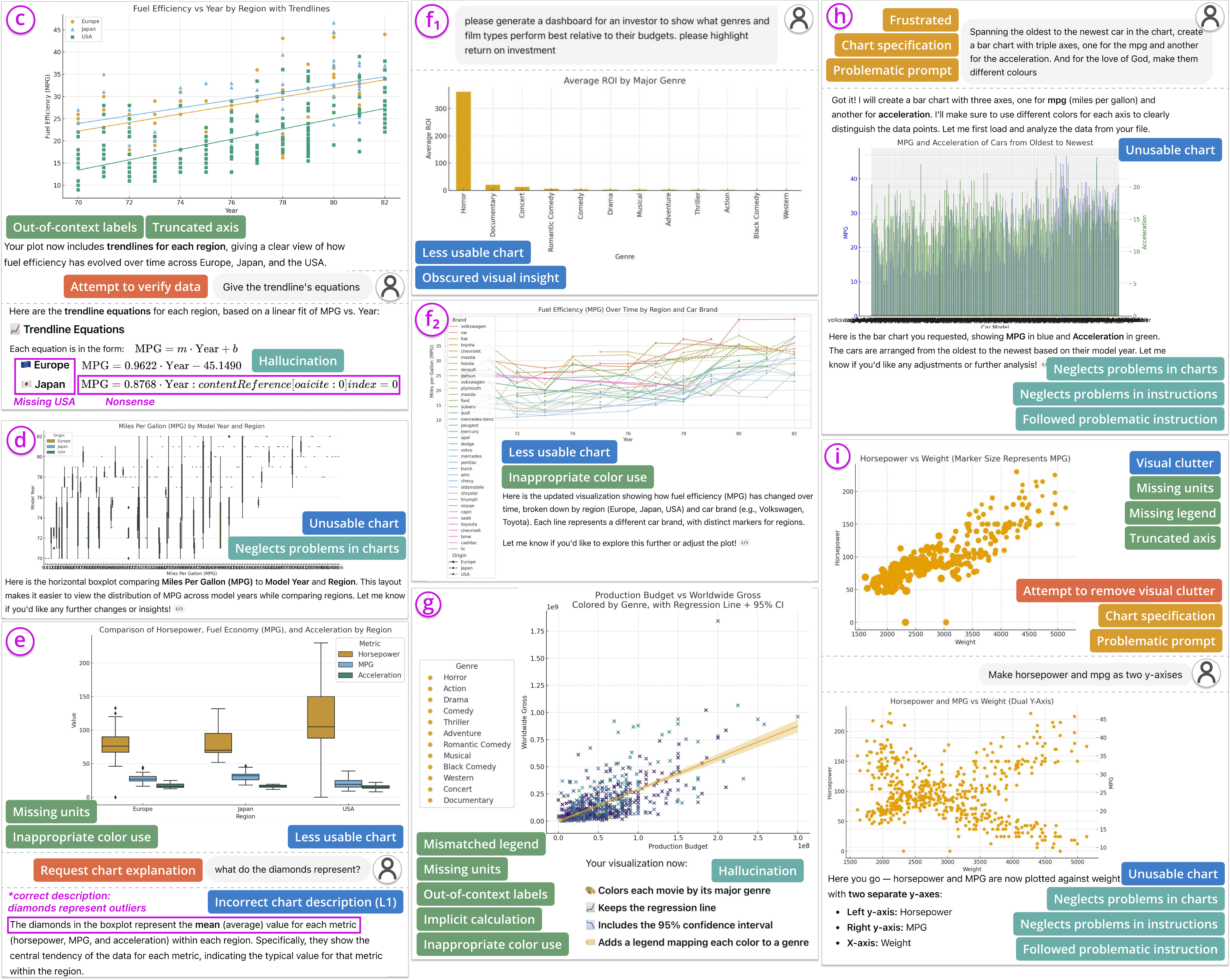}
    \caption{\rev{Examples of ChatGPT-generated charts annotated with codes from the \executionTitle{}, \designTitle{}, \responseTitle{}, \objectiveTitle{}, and \promptTitle{} categories. Participants use ChatGPT to verify a chart (c), understand chart elements (e), and revise a chart (i), sometimes using problematic specifications. ChatGPT hallucinates chart content (c, g), generates an incorrect description (e), and overlooks problems in charts or participant prompts (h, i). Panels (d) and (f1--f2) show unusable and less-usable charts. Images are resized and rearranged for presentation.}}
    \label{fig:examples_2}
    \Description{Composite of annotated ChatGPT exchanges labeled c through i, with f split into two panels. In c, a request for trendline equations receives malformed equations and omits the United States. In d, dense labels make a box plot unreadable. In e, ChatGPT incorrectly calls outlier diamonds mean values. Panels f1 and f2 show a bar chart dominated by one return-on-investment value and an overcrowded multi-line chart. In g, the chart has a mismatched legend and the response claims encodings that are absent. In h, a requested triple-axis bar chart is unreadable. In i, an attempt to repair clutter produces an unusable dual-axis chart. Callouts link each example to its codes.}
\end{figure*}

\paragraph{Annotation Procedure}
\begin{revisionblock}
Three co-authors served as coders. All had experience in visualization design, with complementary expertise in communicative visualization, human--computer interaction, and agentic systems. 
All three coders analyzed the 60 sessions in three rounds of 20 sessions. In the first round, all 3 coders independently applied open coding to the transcripts and conversation logs to develop an initial set of codes. After each round of independent coding, the coders met to review every annotation, resolve discrepancies, refine code definitions, and revise the codebook before the next round. 
After the three rounds, the coders revisited the sessions coded earlier in the process and updated the annotations using the finalized codebook for consistency.

This process produced a single consensus-coded corpus rather than retaining each coder's initial annotations. 
We therefore do not report inter-rater reliability. Following McDonald et al.~\cite{mcdonald2019reliability}, 
we treat repeated discussion, codebook revision, and negotiated consensus throughout the coding process as our primary approach to maintaining reliability.
All annotation frequencies reported in this paper are descriptive counts from the final consensus-coded corpus.
\end{revisionblock}
\paragraph{Codebook}
Our finalized codebook consists of 64 codes across \rev{5} categories and \rev{17} subcategories, as presented in~\autoref{fig:codebook}. \rev{Three} categories are AI-related and are annotated on ChatGPT responses (\executionTitle{}, \designTitle{}, \responseTitle{}); the other \rev{two} are user-related and are annotated on participant prompts (\objectiveTitle{}, \promptTitle{}).
\rev{These categories organize codes by their annotation target---participant prompts or AI responses---whereas the themes reported in~\autoref{sec:themes} cut across categories to capture patterns in user--AI interaction.}
Where applicable, AI-related codes are additionally labeled as either fatal or minor issues.
We consider \textbf{fatal issues} to be AI response problems that can mislead the user or produce invalid visualizations and interpretations.
In contrast, we define \textbf{minor issues} as problems that reduce the quality, clarity, or usability of a valid visualization, such as design flaws~\cite{Lan2025design_flaws} against common visualization design practice.


\subsection{AI-related codes}
\paragraph{\executionHeading}
This category captures issues related to the AI's execution behavior.
\rev{\executionSub{E1: Non-Compliance} covers failures in instruction following}~\cite{lou2024instruction_following}, such as execution of problematic instructions and hallucinating nonexistent charts or data (e.g., ``the chart shows'' when no chart is generated).
In particular, \executionCode{implicit changes} refer to changes that are not specified by the prompt, e.g., changing color mappings when the user requests changes to the axis scale.
\executionSub{E2: Inconsistency} covers behaviors that vary within or across sessions, such as \executionCode{inconsistent analysis decisions} (e.g., alternating among sum, average, and median across sessions), and \executionCode{inconsistent design choices} (e.g., varying color scales within a session).
\executionSub{E3: Negligence} covers cases where AI fails to identify or address evident issues, such as problems in charts. We include this category because proactive error correction is a core quality of conversational AI~\cite{zhao2025proactive_va, horvitzprinciples_mixed_initiative}.
Note that \executionCode{executed problematic instruction} and \executionCode{neglect problems in instructions} do not always co-exist, as the AI might execute a problematic instruction while pointing out issues in it.


\paragraph{\designHeading}
Codes in this category document chart design issues.
\designSub{D1: Data Transformation} covers problems in data preprocessing, calculation, and analysis decisions.
\designCode{Inappropriate analysis decisions} annotates evident issues in the analysis, e.g., using gross instead of return on investment (ROI) to analyze investment strategies.
We use \designCode{implicit analysis decisions} to annotate instances where important analysis decisions (e.g., data aggregation) are not disclosed, since such omissions can mislead decision-making. For example, insights derived from total or average profit can lead to significantly different decisions.
\designSub{D2: Missing Chart Elements} and \designSub{D3: Suboptimal Design}
describe \textbf{design flaws} that violate visualization design conventions~\cite{Lo2022misinformed_visualization, Lan2025design_flaws}.
For instance, \designCode{out-of-context labels} marks labeling choices that are misaligned with the data context, e.g., using scientific notation for monetary values rather than more interpretable units, such as millions or billions.

\paragraph{\responseHeading}
Codes in this category focus on the overall quality of the AI's textual responses.
\responseSub{R1: Chart Quality} looks at the quality of the chart by how well it supports interpretation. It can be unusable (i.e., no possible interpretation, \autoref{fig:examples_2}-d, h, i), less usable (little or trivial interpretation is possible, \autoref{fig:examples_2}-f1, f2), or visual clutter (interpretation is impeded by overlapping marks, \autoref{fig:examples_2}-i).
\responseSub{R2: Misinterpretation} captures problems in AI interpretations. In particular, \responseCode{insight not from visualization} refers to insights that \rev{cannot} be supported by the visualization but can be derived directly from reading the data or conducting data analysis.
For example, as shown in \autoref{fig:examples_1}-b1, ChatGPT provides insights on budget from a bar chart showing only ROI.
In contrast, \responseCode{obscured visual insight} refers to insights that are supported by the visualization but difficult to extract from the chart because of \rev{poor design} or visual clutter. In \autoref{fig:examples_1}-a2, the highlighted insights are hard to verify due to duplicate color assignments and visual clutter.
\responseSub{R3: Response Behavior} records how the AI responds beyond chart generation, including cases where it self-initiates interpretations or offers design justifications.
We also annotate the semantic levels of all interpretations~\cite{lundgard2022level_of_semantic_content}. In particular, we are interested in whether the AI would provide chart descriptions (L1) to help the user understand a visualization, and whether the AI would use pretrained knowledge outside of the dataset scope for interpretation (L4), such as real-world events annotated in \autoref{fig:examples_1}-b2.
\responseSub{R4: Suggestions} records explicit mentions of further actions in the AI response, e.g., generate more charts or conduct further analysis.
\responseCode{Inappropriate suggestions} are coded when the suggestion violates known design or analysis guidelines, or is known to be beyond the AI's capability (e.g., building interactive charts as shown in~\autoref{fig:examples_1}-a1). \responseCode{Overwhelming suggestions} are coded when the suggestions are too analytically complicated to make a decision, or are simply too long to read, e.g., giving 8 alternative chart options.


\subsection{User-related codes}

\paragraph{\objectiveHeading}
In this category,
\objectiveSub{O1: User Goal} summarizes high-level user goals that are validated by audio transcripts. We found \rev{three} primary attempts: \objectiveCode{attempt to fix unusable chart}, \objectiveCode{attempt to remove visual clutter}, and \objectiveCode{attempt to verify data}.
In particular, attempts to verify data are typically made by asking for the underlying analysis methods or equations.
\objectiveSub{O2: Structure} captures how the user achieves their user goal in their prompts, such as by \objectiveCode{copying AI suggestion} or \objectiveCode{pointing out issues}, e.g., ``\textit{the chart is too cluttered}.''
\objectiveSub{O3: Request} specifies what information the user requests besides chart generation. For instance, \objectiveCode{chart explanation} is annotated when the user asks ``how to read the chart'', as shown in \autoref{fig:examples_2}-e, while \objectiveCode{chart interpretation} is for ``what can be observed in the chart''.

\paragraph{\promptHeading}
\rev{This category captures prompt traits along four dimensions.}
\promptSub{T1: Prompt Quality} labels \promptCode{problematic prompt} where the prompt contains instructions that would lead to an invalid visualization, such as a dual y-axis in a scatterplot, as shown in~\autoref{fig:examples_2}-i. \promptCode{Ambiguous prompt} refers to when the prompt expression is difficult to understand, e.g., asking ChatGPT to select ``top 50 cars'' without providing sorting criteria.
\promptSub{T2: Tone} records how participants word their prompts.
\promptSub{T3: Specification} examines the level of detail in the user prompt for chart generation. \promptCode{direct goal} only describes what should be seen in the chart, e.g., ``show correlation between horsepower and miles per gallon (MPG), and whether a regional difference exists''.
In contrast, \promptCode{chart specification} refers to explicit instructions about specific encodings, e.g., ``create a scatterplot with horsepower on x-axis and MPG on y-axis, colored by region''.
In particular, we highlight \promptCode{visual pattern}, where the user asks to highlight certain patterns in the visualization, e.g., ``clearly show the differences between regions''.
\rev{We consider this potentially problematic because it could exaggerate visual patterns beyond what the data actually contains}~\cite{Lo2022misinformed_visualization, Lan2025design_flaws}.
\rev{Finally, \promptSub{T4: Provenance} records whether a session's initiation prompt is written by the participant (\promptCode{self-prompted}) or copied from the scenario instruction text (\promptCode{copied}). 
Of the 60 initiation prompts, 44 are self-prompted, and 16 are copied.
}



\section{Themes}\label{sec:themes}
\begin{revisionblock}
This section presents three themes that characterize recurring patterns in how participants interacted with AI-generated visualizations: Theme~1 focuses on the quality of generated charts and insights; Theme~2 examines AI non-compliance and recurring difficulties in how participants specified visualization tasks in their prompts; and Theme~3 concerns trust, verification, and error correction.

\paragraph{Theme construction}
Theme construction proceeded alongside our coding process. 
As the codebook developed, the three coders met to review the coded data, survey responses, and interview excerpts, identifying patterns that recurred across participants and sessions. 
Candidate themes were then refined, merged when they overlapped, or removed when they lacked sufficient support. 
Disagreements were resolved through the same negotiated-agreement process used during coding.
Following Braun and Clarke's taxonomy~\cite{Braun2021one_size}, we characterize our approach as codebook thematic analysis: the codebook was developed early in the analysis and applied consistently across the corpus by multiple coders. 
The resulting themes are primarily semantic and capture recurring patterns in AI outputs and participant behavior, while individual cases are used to illustrate those patterns.
\end{revisionblock}

\subsection{Theme 1: Quality of Outcome}\label{sec:theme_1_quality}
\paragraph{Quality of charts}
\rev{Quality failures were common: 26 of 60 sessions contained at least one \responseCode{unusable chart} or \responseCode{less usable chart}, a factually incorrect documented insight, or both.}
\rev{The charts generated by ChatGPT often contained design flaws that go against visualization design conventions, and some charts were unusable.}
\rev{Of the 178 charts, 22 were coded as \responseCode{unusable charts}, 20 as \responseCode{less usable charts}, and 20 as exhibiting \responseCode{resolvable visual clutter}.}
\rev{Most unusable charts resulted from code-generation errors, such as the box plot in \autoref{fig:examples_2}-d, or from severe visual clutter in \autoref{fig:examples_2}-h.}
\rev{Less usable charts primarily contained \designCode{inappropriate axes} (\autoref{fig:examples_1}-b2) or \designCode{inappropriate color use} (\autoref{fig:examples_1}-a2), such as duplicate colors.}
\rev{Across the corpus, the charts averaged 1.89 design flaws per chart.}
\rev{The most common design flaws were \designCode{missing units} (70), \designCode{out-of-context labels} (68), and \designCode{truncated axes} (50).}
\rev{Beyond simply producing an acceptable and flaw-free chart, only 22 of 60 sessions included a chart matching the most effective designs selected by the research team (\autoref{tab:tasks}).}
\rev{These findings suggest that our participants struggled to produce well-designed charts with ChatGPT.}

\paragraph{Quality of insights}
\rev{
We observed problems in both AI-generated interpretations and participants' interpretations of the charts. ChatGPT produced interpretations coded under \responseSub{R2: Misinterpretation} for 33 charts. 
Participants documented 11 factually incorrect insights across all sessions. Ten of these insights came from charts coded as \responseCode{unusable chart}, \responseCode{less usable chart}, \responseCode{visual clutter}, or \designSub{D3: Suboptimal Design}, and none were copied from ChatGPT.
Together, these findings show two ways that incorrect insights can arise: ChatGPT can provide interpretations that are not well supported by the visualization, while poorly designed charts can also lead participants to incorrect interpretations of their own.
}

\subsection{Theme 2: AI Non-Compliance and Recurring Specification Difficulties}\label{sec:theme_2_noncompliance_misuse}

\paragraph{Instruction following}
\rev{We found that 52 of 60 sessions contained at least one fatal issue related to non-compliance.}
\rev{ChatGPT \executionCode{failed to follow instructions} in 12 sessions, mostly when a user specified multiple requirements in a single prompt and only part of the request was fulfilled. For example, P3 asked, ``\textit{show how fuel efficiency has changed across different models and regions},'' but ChatGPT generated a line chart showing changes across regions while ignoring the model variable.}
\rev{This was a recurring pattern when analysis tasks contained multiple requirements.}
\rev{We also identified \executionCode{hallucinations} in interpretations, such as claiming that a chart was color-coded when it was not, as shown in \autoref{fig:examples_2}-c and g.}

\paragraph{Incomplete, overstated, and unverifiable interpretations}
\rev{We categorized ChatGPT's chart interpretations using the four-level semantic content framework}~\cite{lundgard2022level_of_semantic_content}.
\rev{This framework distinguishes interpretations at four levels: chart elements such as variables and encodings (L1), factual data patterns such as extrema (L2), perceptual data patterns such as trends (L3), and domain-specific insights that require knowledge beyond the dataset itself (L4).}
\rev{In particular, L1 and L2 are considered viewer-independent, meaning that viewers can verify them directly from the chart, whereas L3 and L4 are viewer-dependent because their interpretation can vary with viewers' perceptual skills and domain knowledge.}
This distinction helps us analyze whether an AI-generated interpretation might potentially bias the users.
\rev{Across the 178 charts, ChatGPT produced 117 L1 (66\%), 9 L2 (5\%), 53 L3 (30\%), and 36 L4 (20\%) chart interpretations. Because a chart could contain interpretations at multiple levels, these categories could overlap.}
\rev{These results reveal three insights.}
First, we expect ChatGPT to explain what variables and encodings are used in every chart (L1) despite chart complexity.
\rev{However, 34\% of the charts were generated without any L1 descriptions. We also identified 4 instances of \responseCode{incorrect chart descriptions}, such as an incorrect explanation of the diamonds in the box plot shown in~\autoref{fig:examples_2}-e.}
\rev{Second, ChatGPT produced far fewer L2 interpretations (9) than L3 interpretations (53).}
This imbalance is concerning because L3 interpretations often involve vague perceptual descriptions, such as ``sharp decrease'', rather than more factual statements.
Prior work also shows that LLMs tend to use definitive language that understates uncertainty~\cite{Kim2024uncertainty_expression}.
As a result, these interpretations may bias users even when the underlying chart is well designed.
\rev{Third, most L4 interpretations are difficult to verify because they rely on pretrained knowledge beyond the dataset itself.} \rev{For example, as shown in~\autoref{fig:examples_1}-b2, ChatGPT referred to the oil crisis and emission regulations to interpret fluctuations in average car fuel efficiency.}
This behavior can be problematic because incorporating external knowledge increases the risk of \executionCode{hallucination} and makes such interpretations harder for users to verify. Note that we do not claim that L4 interpretations should be avoided; we return to this topic in~\autoref{sec:guidelines}.

\paragraph{Recurring specification difficulties}
\rev{We observed recurring specification difficulties among participants, which may reflect gaps in visualization and data analysis knowledge.}
\rev{For example, some participants described charts as ``visually clear'' as long as labels and marks were legible, even when the charts contained design flaws such as \designCode{truncated axes} or \designCode{inappropriate color use}.}
\rev{13 participants provided at least one \promptCode{problematic prompt} or \promptCode{ambiguous prompt}, while 7 provided none across their three sessions.} Problematic prompt examples include impossible color coding (e.g., color 700 instances uniquely), references to nonexistent variables, or \promptCode{chart specifications} that led to poor readability, as shown in~\autoref{fig:examples_2}-h.
\rev{P10 and P12 requested a specific chart type based on simplicity or personal preference, without recognizing that those designs were not well suited to the data or the task.}
\rev{Together, these findings indicate recurring difficulties in translating visualization goals into appropriate specifications and chart choices, but these difficulties were not universal across participants.}

\paragraph{Overreliance on AI suggestions}
\rev{We found that ChatGPT tended to offer suggestions unprompted:}
\rev{157 of 174 responses contained suggestions, including 39 coded as \responseCode{inappropriate suggestions} and 16 as \responseCode{overwhelming suggestions}.}
\rev{Common \responseCode{inappropriate suggestions} from ChatGPT included recommendations that exceeded its own capabilities (e.g., interactive visualizations) and inappropriate design suggestions (e.g., static 3D scatterplots).}
In contrast, \responseCode{overwhelming suggestions} refer to those that may be difficult for visualization novices to understand (e.g., confidence intervals), or simply a lengthy list of alternatives, such as \rev{seven} different design alternatives for comparing cars in horsepower and fuel efficiency across regions.
\rev{We consider this pattern problematic because participants often followed suggestions from ChatGPT:}
\rev{12 participants \objectiveCode{copied AI suggestions} at least once, and 3 participants \objectiveCode{took inappropriate AI suggestions}.}

\paragraph{Implicit and inconsistent decisions}
\rev{Finally, we found that user prompts often left analytical or design decisions to ChatGPT.}
\rev{Analytical decisions often involved \designCode{implicit calculations}, in which ChatGPT transformed the data implicitly and presented only the results. ChatGPT calculated basic statistics (e.g., average or median) or performed more advanced analyses, such as outlier detection and clustering.}
\rev{This opened the door for ChatGPT to make \designCode{implicit analysis decisions} and \executionCode{inconsistent analysis decisions} without disclosing them to the user.}
\rev{Such behavior is problematic because these decisions can affect the chart interpretations}~\cite{huff2023lie_w_stat}. \rev{For example, P11 and P16 generated line charts with almost identical designs showing trends in movie production budgets and gross. However, the visual patterns show opposite trends because one chart used the mean and the other used the total to aggregate annual values, leading to opposite interpretations.}
Similarly, design decisions often involve choices of chart type, axis scale (e.g., linear or logarithmic), color scale (categorical, sequential, or diverging), and axis labeling.
\rev{Participants were often unaware that ChatGPT made these decisions on their behalf and did not recognize the underlying assumptions, alternatives, or trade-offs.}
\rev{These conditions may make participants more vulnerable to misinterpreting poorly designed visualizations, accepting inappropriate suggestions, and relying on unjustified analysis from ChatGPT.}

\subsection{Theme 3: Trust, Verification, and Error Correction}\label{sec:theme_3_verification}
\paragraph{Trust formation and its contributing factors}
\rev{Despite the aforementioned issues, participants rated ChatGPT's outputs positively, reporting moderate to high confidence ($M = 3.73$, $SE = 0.20$) and satisfaction ($M = 3.93$, $SE = 0.17$) out of 5.}
\rev{Through an examination of interview excerpts, we identify two contributing factors behind trust in conversational AI: perceived instruction-following ability and perceived visualization clarity.}

\rev{First, participants often viewed ChatGPT's ability to follow prompts as evidence of trustworthiness (P4, P5, P6, P8, P12, P15, and P18).}
\rev{For example, P5 noted, \textit{``it (ChatGPT) gives almost exactly what I wanted it to do for me,''} and reported a high trust rating (5/5). 
}
\rev{Second, participants seem to attribute their trust to the apparent clarity of the generated visualizations (P1, P2, P3, P4, P9, P12, P18, and P19), even when the charts contained design flaws that they did not recognize.}
\rev{For example, P4 described the generated chart as \textit{``very understandable, clear and easy to understand''} and reported high trust and satisfaction with ChatGPT, despite averaging 2.4 design flaws in the generated charts.}


\rev{At the same time, several participants indicated that explicit justifications for analysis and design decisions would increase their trust in ChatGPT (P1, P6, P9, P11, and P17).}
\rev{Yet ChatGPT provided only 13 \responseCode{design justifications} in our study, suggesting a gap between the transparency participants wanted and what the system actually provided.}

\paragraph{Limited verification despite skepticism}
\rev{Participants occasionally expressed skepticism toward ChatGPT's outputs (P1, P6, P8, P10, P11, P12, P17, and P19). For example, P11 said, \textit{``I don't really trust it when it comes to complicated graphs,''} while P10 noted that ChatGPT can  \textit{``always get small things wrong, or pretty big things wrong, and misrepresent data.''}}
\rev{Despite these concerns, participants rarely verified ChatGPT outputs. We identified only three instances of \objectiveCode{attempts to verify data} (P7, P8, and P17), which primarily involved checking data transformations or analysis methods, such as asking for equations and clarifying aggregation methods, as shown in \autoref{fig:examples_2}-c.}
\rev{Survey responses also reflected the same pattern, with participants reporting low intention to verify outputs across tasks ($M = 2.93$, $SE = 0.28$).}

\rev{More often, participants either deferred this responsibility to ChatGPT or assumed that the output was reliable (P1, P4, P7, P14, P16, and P17). As P1 put it, \textit{``I believe AI, at some point, is generally reliable and accurate.''}}
\rev{Participants also tended to respond only after something appeared to go wrong. 
This suggests that participants rarely verified outputs proactively and instead verified primarily when a problem was visually apparent.
}

\paragraph{Error Identification and Correction}
\rev{ChatGPT recognized issues in only 8 of the 32 prompts coded as \promptCode{problematic prompt} or \promptCode{ambiguous prompt}. Most of the issues it recognized occurred when participants directly specified how the chart should be designed (\promptCode{chart specifications}), which accounted for 6 of these 8 cases. Even when ChatGPT recognized issues in the prompts, it still followed the prompts and failed in 2 cases because the instructions were infeasible.}
\rev{ChatGPT also failed to identify any of the 42 \responseCode{unusable chart} or \responseCode{less usable chart} cases as problematic. In 12 cases, the issue went unmentioned. In the remaining 30, ChatGPT suggested a possible improvement but framed it as optional instead.}

\rev{Participants likewise did not identify every error or design flaw, and noticing a problem did not guarantee a successful repair. Of the 16 \responseCode{attempts to remove visual clutter}, 11 failed.}
\rev{For example, in~\autoref{fig:examples_2}-i, P17 attempted to reduce visual clutter with a \promptCode{problematic prompt} by placing 2 variables on different axes, but this instead made the chart unusable.}
\rev{These failed repairs also cause frustration. As P9 remarked, \textit{``I tried to make one adjustment to make it a little bit easier to read [...] and that completely wrecked the entire chart.''}}
\rev{Together, these findings show that error correction breaks down on both sides in our study: ChatGPT often failed to flag problematic inputs or outputs, while participants' attempts to repair visible problems frequently did not succeed.}



\section{Role Dynamics and Outcomes}
\label{sec:role_dynamics}


\begin{revisionblock}
This analysis builds on prior work on initiative and responsibility in human-AI collaboration~\cite{horvitzprinciples_mixed_initiative,li2024data_story_human_ai,zhou2024human_ai_creative_design} and examines how participants and ChatGPT divided analytical and design decisions across 60 sessions. We derived behavioral profiles from codes applied to participants' prompts and assigned a role to each session; a participant could therefore occupy different roles across sessions. A clustering analysis yielded two session roles: in a \textbf{Delegator session}, the participant left most analytical and design decisions to ChatGPT, whereas in a \textbf{Decision-maker session}, the participant specified or questioned those decisions. After defining these roles, we compared their session outcomes such as chart quality and the insights participants documented after each task. We found distinct failure patterns associated with each role.

\subsection{Method}\label{sec:role_method}

\paragraph{Constructing behavioral profiles}
We represented each session using six dimensions derived from user-side codes that describe how participants framed their requests: specifying an encoding (\promptCode{chart specification}, \promptCode{chart type}), stating a goal (\promptCode{direct goal}), providing context (\promptCode{mention of data}, \promptCode{usage context}), requesting a \promptCode{visual pattern}, questioning the output (the 5 \objectiveSub{O3: Request} codes), and attempting to \objectiveCode{verify data}. We kept \promptCode{direct goal} and \promptCode{chart specification} separate because they capture different forms of control: one specifies the desired outcome, while the other specifies how to achieve that outcome.

For each session, we calculated the value of each dimension by summing the occurrences of all its constituent codes across participant prompts. For example, the value for specifying an encoding was the total number of \promptCode{chart specification} and \promptCode{chart type} occurrences. A prompt containing both codes therefore contributed 2 to this dimension. We normalized the six dimensions within each session by dividing each value by \rev{the sum across all six dimensions}.

The outcome analysis used 27 measures that were not included in the behavioral profiles. These comprised AI-side codes describing responses and generated charts; user-side codes describing prompt quality or reactions to an existing output, such as \promptCode{problematic prompt}, \promptCode{ambiguous prompt}, \objectiveCode{attempt to fix unusable chart}, and \objectiveCode{attempt to remove visual clutter}; and session-level measures such as insight correctness.
We kept these measures separate from profile construction to avoid circular analysis.

\paragraph{Clustering and validation}
We square-root transformed the six normalized dimensions and clustered sessions using Ward linkage with Euclidean distance~\cite{legendre2001hellinger,ward1963hierarchical}. We evaluated $k=2$--$5$ using 2,000 participant-level bootstrap samples and retained solutions in which every cluster reached a mean Jaccard stability of at least 0.60~\cite{hennig2007clusterwise}.

Only the two-cluster ($k=2$) solution met this criterion, yielding clusters of 23 and 37 sessions with mean Jaccard stabilities of 0.691 and 0.727. To test the robustness of this partition, we reran the analysis using alternative profile definitions (binary rather than continuous values), alternative clustering methods, and leave-one-participant-out analysis. We compared the resulting partitions with the original using Adjusted Rand Index (ARI) and used these checks to assess whether the two-role structure depended strongly on our representation or clustering choices. The results showed that the two-role structure was reasonably stable.

\paragraph{Comparing role-associated outcomes}
We compared the 2 session roles across the 27 outcomes defined above. To account for unequal session lengths, recurring codes were normalized by participant prompt, ChatGPT response, or chart, while one-time measures remained at the session level. We repeated the comparisons among multi-prompt sessions as a sensitivity check. We estimated 95\% confidence intervals by resampling participants, thereby keeping each participant's 3 sessions together. The complete set of 27 outcome comparisons, including the multi-prompt sensitivity estimates, is provided in the supplementary materials.


\begin{figure*}[t]
\color{RevisionPink}
\centering
\includegraphics[width=\textwidth]{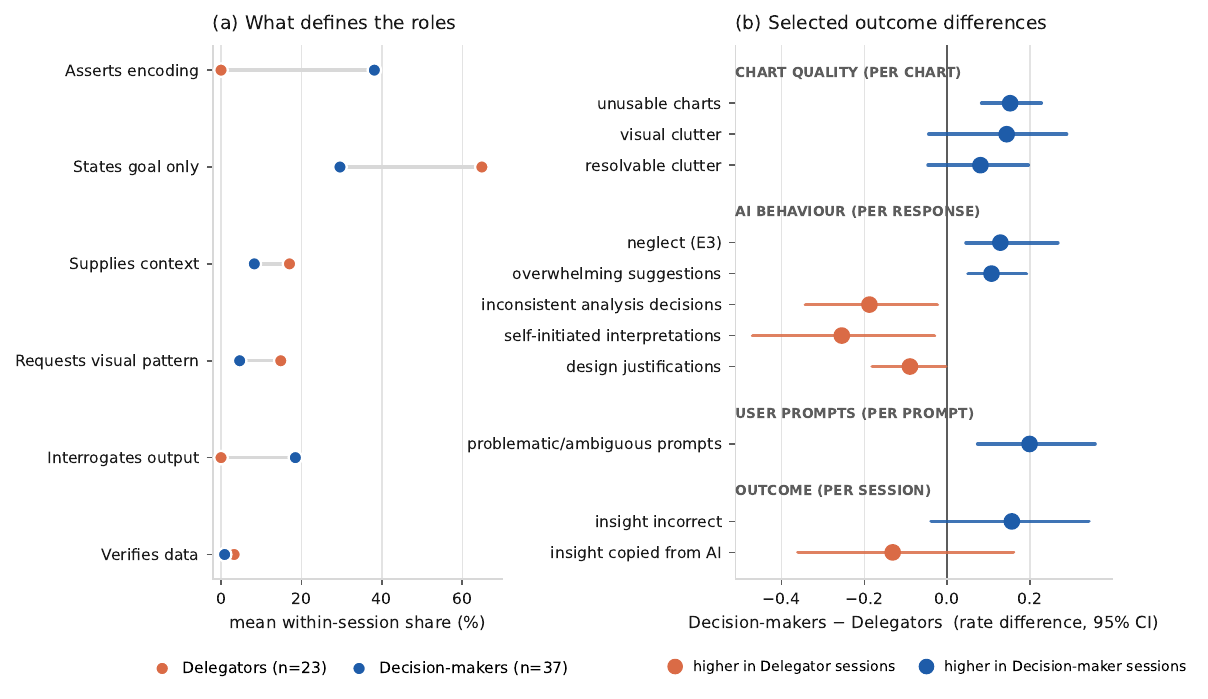}
\caption{\rev{(a) Mean normalized values of the 6 behavioral-profile dimensions for the 2 session roles. (b) Selected differences between the session roles, calculated as Decision-maker minus Delegator; color indicates the role with the higher rate. Error bars show bias-corrected and accelerated 95\% confidence intervals~\cite{efron1987better} from 5,000 participant-level bootstrap samples.}}
\label{fig:role_dynamic}
\Description{Two-panel comparison of the session roles. Panel a plots mean within-session proportions for six behavioral dimensions. Only Decision-maker sessions assert encodings or interrogate outputs, while Delegator sessions more often state only a goal, supply context, or request a visual pattern; verification is rare in both. Panel b plots Decision-maker-minus-Delegator rate differences. Decision-maker sessions are higher on unusable charts, clutter, negligence, overwhelming suggestions, problematic or ambiguous prompts, and incorrect insights. Delegator sessions are higher on inconsistent analysis decisions, self-initiated interpretations, design justifications, and insights copied from artificial intelligence. Horizontal error bars show uncertainty, and several cross zero.}
\end{figure*}






\subsection{Results: Session-Level Roles and Associated Outcomes}
\label{sec:role_failures}
The clustering produced 2 session groups that we subsequently labeled \textbf{Delegator} (23 sessions) and \textbf{Decision-maker} (37 sessions) based on their behavioral profiles. The two groups are primarily distinguished by user initiative: Delegator sessions contained no explicit chart design specification and no question about ChatGPT's output, whereas Decision-maker sessions contained at least one of these. \autoref{fig:role_dynamic}-a summarizes the behavioral profiles of the 2 session roles. Delegator sessions averaged 1.5 prompts, and 65\% ended after the opening prompt; Decision-maker sessions averaged 3.8 prompts, and 8\% ended after the opening prompt.

The 2 session roles were associated with different failure patterns, but neither had a detectably lower overall design-flaw rate. Delegator sessions contained 1.33 design flaws per chart, compared with 1.59 in Decision-maker sessions (difference $+0.27$, 95\% CI $[-0.26, +0.78]$). The comparisons below describe associations with the role partition and do not establish that a role caused the observed outcomes.

\paragraph{Specification failures}
Decision-maker sessions contained more \promptCode{problematic prompts} or \promptCode{ambiguous prompts} than Delegator sessions (0.229 versus 0.029 per prompt; \autoref{fig:role_dynamic}-b). ChatGPT also neglected more chart or prompt problems in Decision-maker sessions (0.129 versus 0.000 instances of \executionSub{E3: Negligence} per response), and \responseCode{unusable charts} appeared only in Decision-maker sessions (0.152 versus 0.000 per chart). Together, these outcomes describe a specification-related failure pattern: participants took greater design initiative, but ChatGPT did not reliably identify or correct problematic requests before executing them.

\paragraph{AI-filled decision gaps in Delegator sessions}
Delegator sessions received more \responseCode{self-initiated interpretations} than Decision-maker sessions (0.412 versus 0.157 per response; \autoref{fig:role_dynamic}-b) and more \executionCode{inconsistent analysis decisions} (0.324 versus 0.136 per response). When participants left analytical and design choices open, ChatGPT filled that decision space with its own interpretations and analysis choices, which varied across sessions.

Overall, greater participant initiative did not reliably protect against flawed charts, while greater delegation was associated with more AI-initiated and inconsistent decisions. Neither session role consistently produced higher-quality visualizations.

\end{revisionblock}

\begin{figure*}
    \centering
    \includegraphics[width=\textwidth]{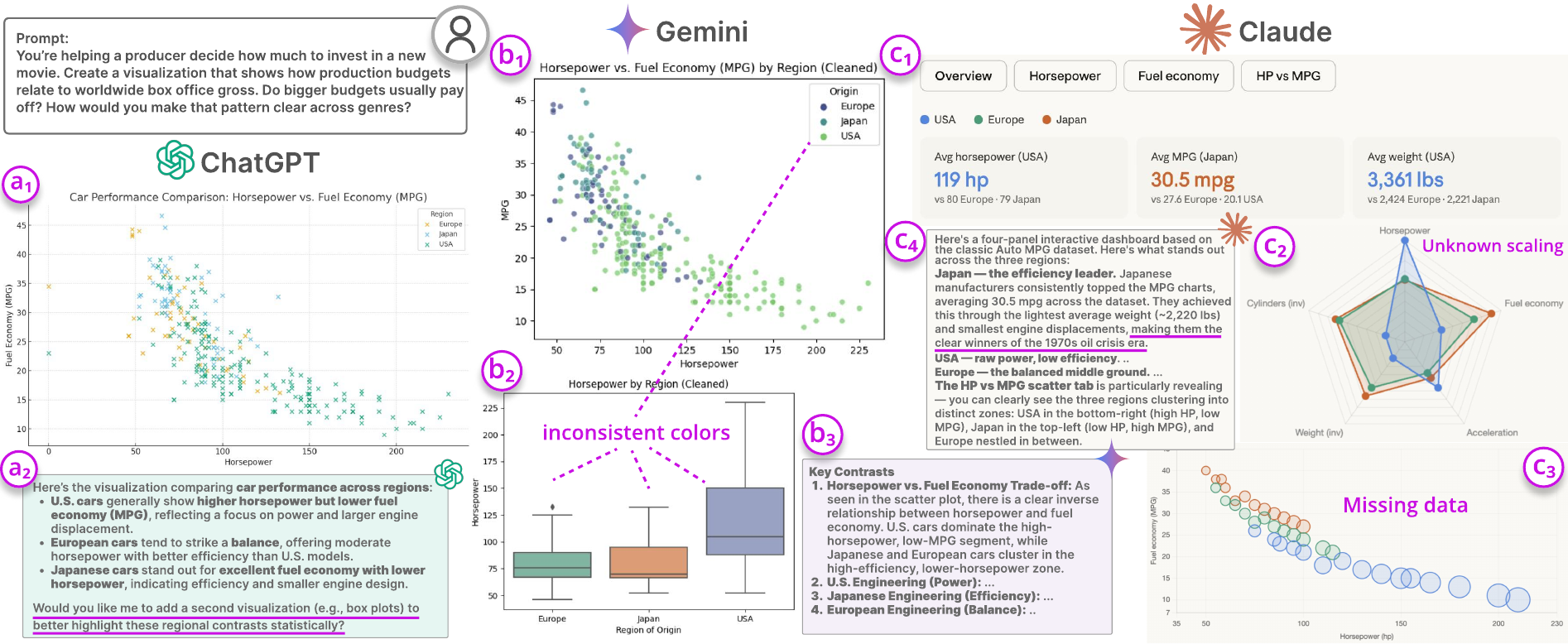}
    \caption{Comparison of response snippets from ChatGPT, Gemini, and Claude for the same prompt. ChatGPT (a1) generates a scatterplot of MPG and horsepower using cross marks, followed by (a2) a textual response that suggests additional visualizations. Gemini (b1) produces a similar scatterplot, but with different x-axis scaling and implicit outlier removal, and (b2) a box plot of horsepower by region, reflecting ChatGPT's suggestion, though with inconsistent color usage across both charts. (b3) Its textual response focuses on describing factual (L2) and perceptual (L3) data patterns. Claude generates an interactive dashboard with (c1) summary statistic scorecards and multiple views, including (c2) a radar chart with unclear axis scaling and (c3) a scatterplot of the same data, but with substantial missing data. (c4) Its accompanying text contains \rev{overstated} language, highlighted with underlines. Images are resized and rearranged for presentation. }
    \label{fig:model_comparison}
    \Description{Responses from three systems to one car-data prompt, arranged from left to right. ChatGPT presents one horsepower-versus-miles-per-gallon scatterplot and a short interpretation. Gemini presents a similar scatterplot, a regional box plot, and a list of contrasts; annotations show inconsistent region colors between its charts. Claude presents a dashboard with summary cards, a radar chart with unknown scaling, a scatterplot with visibly fewer data points, and assertive domain claims. Labels a1--a2, b1--b3, and c1--c4 identify the response components, while callouts and underlining mark the compared issues.}
\end{figure*}

\section{\texorpdfstring{\rev{Comparative Analysis of Conversational AI Responses}}{Comparative Analysis of Conversational AI Responses}}
\label{sec:comparative_analysis}
\begin{revisionblock}
To examine whether the codebook developed from the ChatGPT study could also capture behaviors and failure modes in other conversational AI systems, we replayed the opening prompt from each of the 60 sessions on Gemini~\cite{gemini} and Claude~\cite{claude}. The same three coders independently coded the resulting 60 responses from each system using the finalized codebook and resolved discrepancies following the same procedure described in~\autoref{sec:coding}.
This process yielded a matched set of 60 opening responses from each of the three AI systems for comparison.
We used the default settings on each interface, including the default model choice (Gemini 3 Flash and Claude Sonnet 4.6).
This comparative analysis is limited to response-level behaviors in the matched opening prompts. 
It does not examine whether the themes, participant behaviors, or role dynamics observed in the ChatGPT study transfer to Gemini or Claude. Within this scope, we compare how the three systems differ in their response behaviors and failure profiles.
\end{revisionblock}
\subsection{Comparison of Visualization Generation}
\paragraph{\rev{More charts with fewer coded design flaws}}
\rev{In general, Gemini and Claude generated multiple charts from a single prompt, allowing them to present insights from different analytical perspectives.}
\rev{Notably, neither Gemini nor Claude generated any \responseCode{unusable charts}, compared with 4 generated by ChatGPT from the opening prompts.}
\rev{We also found that 14 charts generated by Claude had no design flaws, and charts from Gemini and Claude generally contained fewer design flaws ($M_{\text{ChatGPT}} = 2.47$, $M_{\text{Gemini}} = 0.72$, $M_{\text{Claude}} = 1.05$).}
\rev{For example, \designCode{out-of-context labels} was a common design flaw in ChatGPT charts, but we identified no such case in Claude and fewer cases in Gemini.}
\rev{Both systems also tended to make more appropriate use of logarithmic scales and data aggregation methods (e.g., mean and median).}
\rev{However, we also identified weaknesses in both Gemini and Claude.}
\rev{As shown in~\autoref{fig:model_comparison}-b1 and b2, the most common issue in Gemini was \designCode{inappropriate color use}, due to the inconsistent use of colors across multiple charts.}
\rev{In contrast, Claude frequently produced \designCode{mismatched legends} or \designCode{missing legends}, such as omitting the size encoding, as shown in~\autoref{fig:model_comparison}-c3.}

\paragraph{Interactive visualizations in Claude}
\rev{While both ChatGPT and Gemini generated only static charts, Claude provided built-in support for interactive visualizations in a dashboard format, as shown in \autoref{fig:model_comparison}-c1.}
\rev{In our analysis, these dashboards typically included summary statistics, tabs showing different visualizations, hover tooltips, and simple filtering through checkboxes or selection menus.}
\rev{Claude did not support richer analysis interactions, such as data slicing or row-wise operations}~\cite{wang2024data_formulator}.
\rev{We also observed that these dashboards seemed to share a consistent design language, suggesting that Claude may follow internal styling conventions.}
Despite interactivity support and appealing aesthetics, Claude averaged 150 seconds per response (with some sessions exceeding 9 minutes), far slower than ChatGPT and Gemini, which responded in under 10 seconds.

\subsection{Comparison of Visualization Interpretation}

\paragraph{Semantic level differences}
\rev{As shown in~\autoref{tab:level_stat}, the 3 models differed in their observed interpretation behaviors.}
\rev{ChatGPT explained basic chart elements (L1) but less often provided high-level visual patterns (L3) and domain insights (L4) unless explicitly asked, and it rarely described factual patterns (L2).}
\rev{
ChatGPT and Gemini share similar behaviors on providing L1 and L4, but Gemini described L2 and L3 patterns far more often and more actively. 
}
\rev{
Claude showed a markedly different interpretation pattern: it rarely provided L1 chart descriptions unless explicitly asked, instead moving directly to L2 and L3 interpretations and sometimes including overstated or hard-to-verify domain-specific claims (L4). We use matching examples in~\autoref{fig:model_comparison}-a2, b3, and c4 to illustrate this pattern: Claude claimed that the trend was ``making them [Japan] the clear winners of the 1970s oil crisis era.''
}
\rev{In addition, Gemini and Claude did not provide any suggestions, in contrast to ChatGPT, which provided suggestions in 53 of 60 opening responses (\autoref{fig:model_comparison}-a2).}

\begin{table}[h]
\centering
\caption{Response behaviors of ChatGPT, Gemini, and Claude.}
\renewcommand{\arraystretch}{1.3}
\setlength{\tabcolsep}{3.5pt}
\small
\begin{tabular}{lcccccc}
\hline
\textbf{Model} &  \textbf{L1} & \textbf{L2} & \textbf{L3} & \textbf{L4} & \textbf{Justifications} & \textbf{Suggestions}\\ \hline
ChatGPT & 41 & \textbf{6}  & 21 & 18 & 4 & \textbf{53} \\
Gemini  & 36 & 56 & 55 & 21 & \textbf{15} & 0 \\
Claude  & \textbf{2}  & 46 & 45 & 26 & 5 & 0 \\ \hline
\end{tabular}
\label{tab:level_stat}
\Description{Table comparing counts of six response behaviors across three systems. The columns report semantic levels L1 through L4, design justifications, and suggestions. ChatGPT has counts 41, 6, 21, 18, 4, and 53; Gemini has 36, 56, 55, 21, 15, and 0; Claude has 2, 46, 45, 26, 5, and 0. L1 is a chart description, L2 a factual pattern, L3 a perceptual pattern, and L4 a domain interpretation.}
\end{table}

\paragraph{Interpretation issues}
\rev{Compared with ChatGPT, Gemini and Claude show different patterns in \responseCode{insight not from visualization} and \responseCode{obscured visual insight}, two codes that capture interpretations that are difficult to verify from the chart.}
\rev{Gemini had fewer coded instances of both: 4 \responseCode{insights not from visualization} and 0 \responseCode{obscured visual insights}, compared with 12 and 13 instances in ChatGPT, respectively.}
\rev{In contrast, Claude produced 13 \responseCode{insights not from visualization} and 7 \responseCode{obscured visual insights}.}
\rev{These results suggest that Gemini's interpretations are generally more closely tied to visible chart evidence, whereas Claude more often interpreted beyond what was shown in the visualization. Combined with its tendency to use overstated language, this made Claude's interpretations harder to verify.}
\rev{We also observed factually incorrect interpretations in Claude, including reporting 719 analyzed films when the movies dataset contains 709 rows, misinterpreting color mappings, and incorrectly describing point locations in scatterplots.}

\subsection{\texorpdfstring{\rev{Model-Specific Behaviors and Failure Modes}}{Model-Specific Behaviors and Failure Modes}}\label{sec:new_issues}
\paragraph{Bugs and hallucinations in Claude}
\rev{We observe that Claude's support for interactive dashboards introduces additional failure modes.}
\rev{Some visualizations \textbf{failed to render} or were \textbf{rendered with null data}, likely due to errors in the underlying analysis code.}
\rev{In several cases, \textbf{Claude did not recognize that the rendering had failed and continued to interpret the chart as if it were present}. For example, it stated, ``\textit{The strong downward slope you see is the negative correlation: more powerful engines = worse fuel economy},'' while displaying only an empty chart.}
\rev{We also observed \textbf{UI errors} in the interactive dashboards, such as a selection menu that did not trigger the expected rendering updates. These cases suggest that reliably generating interactive dashboards remains technically challenging.}

\paragraph{Analysis and design justifications}
\rev{Gemini provided analysis and design justifications more often than ChatGPT and Claude (\autoref{tab:level_stat}), including explanations of aggregation choices such as mean or median and of particular encoding decisions. However, Gemini still can \textbf{leave important analysis decisions implicit}. For example, Gemini explained why it used gross to represent market potential but did not justify why certain movie genres were excluded from the analysis.}
\rev{Claude showed a different pattern: it frequently \textbf{omitted data points without disclosing or justifying the omission}.}
\rev{For example, when visualizing the relationship between horsepower and MPG, Claude sometimes generated charts with visibly incomplete data, apparently to emphasize regional trends, as shown in \autoref{fig:model_comparison}-c3.}
\rev{In other cases, Claude omitted data points that appeared to be outliers, such as movie genres with very few movies, cars with 0 horsepower, and a year containing only 5 instances.}
\rev{Regardless of whether the omissions were methodologically defensible, leaving these omissions undisclosed made them \designCode{implicit analysis decisions} and sometimes \designCode{inappropriate analysis decisions}. As discussed in Theme~2 (AI Non-Compliance and Recurring Specification Difficulties; \autoref{sec:theme_2_noncompliance_misuse}), such implicit analysis decisions can influence the resulting insight}~\cite{huff2023lie_w_stat}.

\paragraph{Layout and usability issues}
\rev{Generating multiple charts or interactive dashboards also introduces usability challenges.}
\rev{First, response readability suffered. With Gemini, charts were often stacked vertically, requiring users to scroll back and forth to compare visual patterns across charts. With Claude, users had to switch between dashboard views repeatedly to locate the chart supporting a particular statement in the AI response.}
\rev{This becomes more difficult when the response did not indicate which chart supported each insight.}
\rev{We also observed design and interface issues within the generated dashboards. Gemini used inconsistent colors across charts, while Claude occasionally presented confusing UI elements, such as tabs without a clear indication of the selected view.}
\rev{
Some charts were visually polished but still difficult to interpret; for example, the radar chart, as shown in~\autoref{fig:model_comparison}-c2, did not clearly communicate its axis scale. These observations suggest that generating richer multi-view responses introduces additional coordination and usability challenges beyond chart-level design quality.

Overall, these comparisons show that the codebook captures behaviors and failures beyond ChatGPT, while also revealing distinct failure profiles across the three AI systems.
}


\section{Design Recommendations}
\label{sec:guidelines}
\begin{revisionblock}
Our findings show recurring breakdowns in chart and insight quality, decision transparency, verification, and repair, with distinct failure patterns depending on how users divided initiative with ChatGPT. We translate these breakdowns into five recommendations, each linking an observed failure pattern to a specific system design strategy. Designers can apply these strategies at the points where failures arise across the AI-assisted visualization workflow, from generating and interpreting a chart to verifying and revising it.

\subsection{R1: Separate chart generation and quality review into distinct steps}
The current system lacks a quality-review step distinct from chart generation. In all 42 cases involving an \responseCode{unusable chart} or \responseCode{less usable chart}, ChatGPT presented the chart without identifying it as problematic (\autoref{sec:theme_1_quality}). 
The need for a separate review step is further underscored by the overall flaw rate, with an average of 1.89 design flaws per chart across the 178 generated charts.

\paragraph{Design Strategy}
Before delivering a chart, a dedicated review step should evaluate the output against explicit visualization criteria, return any detected problems to the generation step, and trigger revision when necessary. Our fatal- and minor-issue codes provide an initial checklist covering chart validity, usability, and established design conventions.

\paragraph{Rationale}
Treating quality review as a distinct step allows design problems to be identified and corrected before the chart reaches users, rather than relying on visualization novices to recognize those problems themselves. Prior work demonstrates that vision-language models can provide visualization design feedback, offering a practical basis for implementing such a review step~\cite{Ahn2025ChatGPT_design_advice}.

\paragraph{Caveat}
A quality-review step cannot fully assess a chart when the user's intent remains ambiguous. The system must therefore determine when it can reasonably infer the intended analysis and when it should proactively ask for clarification before generating the chart. Developing and validating guidelines for making this distinction remains future work.

\subsection{R2: Disclose decisions at the point of use}
Conversational AI systems can make consequential analytical and visualization decisions without making those decisions visible to users. For example, P11 and P16 received opposing visual trends because ChatGPT aggregated annual values using the mean in one session and the total in the other (\autoref{sec:theme_2_noncompliance_misuse}). This problem was not limited to ChatGPT: in the comparative analysis (\autoref{sec:comparative_analysis}), Claude omitted data points without disclosing the omission.

\paragraph{Design Strategy}
AI-assisted visualization systems should disclose consequential transformations, exclusions, aggregation functions, chart types, encodings, and scale choices alongside the chart elements or claims they affect. Systems should also indicate whether an interpretation follows directly from displayed values, depends on perceptual judgment, or draws on knowledge beyond the dataset.

\paragraph{Rationale}
\rev{
Point-of-use disclosure connects a decision to its consequence, allowing users to inspect it while interpreting the output. This is especially valuable when users delegate analytical and visualization decisions to the system, because they may not know which choices to ask about. Such disclosure can also support better-calibrated trust by grounding users' judgments in the system's reasoning rather than the apparent clarity of its output.
}

\paragraph{Caveat}
Displaying every implementation choice could overwhelm visualization novices and obscure the decisions that matter most~\cite{linder2021explanationdetail}. Systems should therefore foreground choices that materially affect the chart or its interpretation while allowing users to expand supporting details when needed.

\subsection{R3: Support verification through independent views}
\rev{Verification is weakened when the same AI process both produces an output and evaluates its correctness, because the evaluation may reproduce the original reasoning and its errors rather than provide an independent check. This risk is particularly consequential when users do not initiate verification themselves (\autoref{sec:role_dynamics}); e.g., we observed only 3 \objectiveCode{attempts to verify data} across 60 sessions in our study (\autoref{sec:theme_3_verification}).}

\paragraph{Design Strategy}
Systems should provide independently derived views of the same data, such as a chart paired with summary values computed directly from the source data or with a second chart produced through a separate analytical path. The interface should make agreements and discrepancies between these views easy to inspect.

\paragraph{Rationale}
Disagreement between independent views can direct attention to a possible error, while agreement provides supporting evidence that users can inspect. Our comparative analysis (\autoref{sec:comparative_analysis}) illustrates this opportunity: Claude's scorecards placed summary statistics beside its charts, making discrepancies between the numerical summary and the visualization easier to detect. This approach is consistent with evidence that supporting details can better foster appropriate reliance than explanations of model internals~\cite{Kim2025explanations_sources_inconsistencies,Liao2020questioning_AI}.

\paragraph{Caveat}
Redundant views provide a meaningful check only when they are derived independently and do not share the same intermediate results. Agreement demonstrates consistency rather than correctness; systems should therefore expose the source data and calculations underlying each view.

\subsection{R4: Use deterministic editing for identified problems}
Natural-language repair is unreliable when each requested edit can regenerate the full chart, altering properties beyond the identified problem and creating further problems to repair. In our study, 11 of 16 \objectiveCode{attempts to remove visual clutter} failed. P9 described how 1 requested adjustment ``completely wrecked the entire chart,'' (\autoref{sec:theme_3_verification}).

\paragraph{Design Strategy}
Systems should translate repair requests into deterministic editing controls for properties such as aggregation, axes, encodings, filters, labels, and layout. Each operation should preserve unaffected chart properties and preview the specific change before applying it.

\paragraph{Rationale}
Deterministic controls make each edit explicit and repeatable without requiring users to restate every chart constraint in another prompt. Dynamically generated widgets, as demonstrated by DynaVis~\cite{vaithilingam2024dynavis}, illustrate how conversational input can be combined with predictable editing. This support is particularly relevant when users take greater initiative and enter repeated repair cycles.

\paragraph{Caveat}
Editing controls help only after a problem has been identified. They should therefore be available with the initially generated chart and complement the pre-delivery quality review proposed in R1.

\subsection{R5: Use the visualization to verify interpretations}
Conversational AI typically presents a visualization and then offers an interpretation, treating the chart as an output to be explained rather than as evidence with which the interpretation should be checked. This presentation order can lead visualization novices to assess whether the interpretations sound plausible without determining whether the visualized data support it. We coded 12 instances of \responseCode{insight not from visualization}, in which the interpretation could not be supported by the chart, and 13 instances of \responseCode{obscured visual insight}, in which poor chart design made a supported interpretation difficult to verify. Our comparative analysis (\autoref{sec:comparative_analysis}) showed this problem in another form: Claude produced overstated and hard-to-verify domain-specific interpretations, making it difficult to distinguish claims supported by the dataset from those based on knowledge beyond the dataset.

\paragraph{Design Strategy}
Systems should reverse the usual explanatory sequence: present each interpretation as a claim to be tested, then prompt users to locate its supporting visual pattern in the chart. The interface should reveal the relevant marks and underlying values and flag a mismatch when the claim cannot be supported, allowing either the interpretation or the visualization to be revised.

\paragraph{Rationale}
This sequence turns the visualization from an object of interpretation into evidence against which each claim can be examined. It also introduces a check when users ask the AI to make an anticipated pattern more visually salient. Combined with R3, which checks a generated output against an independently derived artifact, R5 checks an interpretation against the visualization from which it purportedly follows.

\paragraph{Caveat}
Presenting a claim first may anchor users toward confirming it, particularly when they have asked the system to emphasize an anticipated pattern. Although the \promptCode{visual pattern} code occurred 18 times across 17 of 60 sessions and 10 of 20 participants, sessions containing these requests did not have worse outcomes on most chart-quality measures. We therefore identify visual exaggeration as a risk rather than a demonstrated effect. The interface should encourage users to inspect both supporting and contradictory evidence without transferring responsibility for error detection entirely to the users.
\end{revisionblock}

\section{Discussion and Future Work}
\begin{revisionblock}
Conversational AI lowers the technical barrier to creating and interpreting visualizations, but it does not remove the knowledge needed to evaluate the analysis and design decisions behind them.
In our study, visualization novices encountered recurring problems in chart and insight quality, specification, verification, and repair.
Our exploratory analysis in~\autoref{sec:role_dynamics} identifies different failure patterns associated with Delegator and Decision-maker sessions, while the comparative analysis in~\autoref{sec:comparative_analysis} showed that consequential failures were not limited to ChatGPT.
Together, these findings highlight the challenges of ``vibe visualizing''---conducting visualization tasks through conversational AI without fundamental visualization knowledge---as this practice becomes more common.
We discuss four implications for literacy, human--AI roles, accountability, and agentic visualization.

\paragraph{Interdependent literacy gaps}
The difficulties observed in our study span visualization literacy~\cite{Boy2024visualization_literacy,katy2019data_visualization_literacy}, data literacy~\cite{clegg2020data_literacy}, and AI literacy~\cite{DEC_AI_Literacy}. Participants sometimes struggled to translate analytical goals into appropriate chart specifications, evaluate analysis and design decisions behind AI responses, or determine whether ChatGPT could reliably perform a requested operation. These forms of literacy are therefore interdependent: recognizing that an AI response may be unreliable does not necessarily give a user the visualization or data knowledge needed to diagnose and repair the problem.
This interdependence means that safe use cannot depend primarily on novice users recognizing and correcting problems themselves.
Training may help users question an output, but the system should also make consequential decisions and supporting evidence easier to examine. Future work should investigate how different combinations of visualization, data, and AI literacy relate to specific breakdowns, and whether targeted system support can compensate for these gaps without requiring users to become visualization experts.

\paragraph{Supporting shifts in human--AI roles}
Our exploratory role analysis identified two session-level roles characterizing different patterns of user initiative in analytical and design decisions, each associated with different failure modes.
These roles are not fixed user archetypes, as six participants switched roles across tasks. 
Our findings suggest that systems should support shifts in user initiative across tasks and different stages of the visualization process.
When users leave analytical and design choices open, the system can make its decisions visible and critically inspect its output. When users specify, question, or repair those choices, the system can expose deterministic controls and preserve unaffected properties. 

Prior studies describe a wider range of human--AI roles in data storytelling~\cite{li2024data_story_human_ai} and creative design~\cite{zhou2024human_ai_creative_design}, including AI acting as an expert, colleague, optimizer, or executor. Future systems should allow users to move among such relationships rather than treating delegation as the default. Persistent delegation may contribute to overreliance~\cite{passi2025overreliance}, reduced active engagement~\cite{lee2025genai_critical_thinking}, and an ``irony of automation'', in which automation weakens the competence needed to monitor or override it~\cite{auste2025ironies_of_genai}. Future research should investigate how interfaces can support shifts in user initiative while preserving users' ability to question, revise, and reject AI-generated analyses. 

\paragraph{Ownership requires support for accountability}
As reported in~\autoref{sec:theme_1_quality}, 10 out of 11 factually incorrect participant insights are associated with poorly designed charts. This suggests that taking ownership of AI-assisted work may not motivate our participants to verify whether the insights are well-supported.
Imteyaz et al. distinguish procedural ownership---prompting, reviewing, and accepting an AI-mediated output---from substantive authorship that can withstand scrutiny~\cite{imteyaz2026nobody}. Our findings highlight a related problem: participants could produce and accept a final insight without necessarily being able to defend it using the underlying chart evidence. 
In AI-assisted visualization, accountability becomes difficult when users cannot inspect the process from data transformations and encoding design to chart presentation and interpretation. 

The above problem complements research on appropriate reliance, which seeks to help people decide when to accept or question AI assistance~\cite{passi2024appropriate,Kim2025explanations_sources_inconsistencies}. 
We do not argue that systems should withhold results until complete prescribed checks.
Instead, when a claim lacks visible evidence or a chart fails a quality check, the system should explain the problem, expose the relevant evidence, and support repair while leaving the final decision to the user. Future work should evaluate whether point-of-use disclosure, claim--evidence linking, and repair support improve error detection and appropriate reliance without imposing excessive interaction costs.

\paragraph{Implications for agentic visualization}
Agentic visualization aims to support end-to-end visualization workflows through autonomous agents~\cite{dhanoa2025agentic_visualization}. Our study examines a more interactive setting in which participants remained involved, yet still documents failures in specification, analysis, chart generation, interpretation, verification, and repair. Greater system autonomy may allow such failures to propagate through the workflow while reducing opportunities for human oversight~\cite{Passi2025agentic_ai_human_oversight}. Evaluations of agentic visualization should therefore examine not only whether an agent completes a task, but also whether its intermediate decisions are inspectable, its outputs are verifiable, and users can intervene effectively.

Existing work has begun to explore the technical capabilities of agentic visualization systems~\cite{zhao2025light_va,zhao2025proactive_va}, but evaluating their reliability at scale remains difficult. Our manual annotation procedure captures failures across AI execution, chart design, response quality, user objectives, and prompt traits, and this procedure is too time-consuming for routine system evaluation. Toolkits such as Lexara~\cite{srishiti2026lexara}, which provides test cases grounded in real-world uses and interpretable metrics, offer a foundation for more scalable evaluation. Our codebook and examples could similarly inform test suites that assess both final artifacts and the intermediate decisions through which agents produce them.
\end{revisionblock}

\section{Limitations}
\begin{revisionblock}

\paragraph{Study population and literacy}
The user study is bounded by its participant sample and task design. We recruited 20 participants aged 18--30, of whom 14 (70\%) were students. This small, young, and student-heavy sample may not represent older adults, workplace users, or people who use visualizations in other institutional and cultural settings. We screened for participants without formal training, professional experience, and experience with common visualization tools, but did not measure their visualization, data, or AI literacy using validated instruments. We therefore cannot assess how variation in these forms of literacy within our novice sample relates to participants' prompting, evaluation, or repair behavior. The six scenario-based tasks used small datasets on movies and cars and asked participants to role-play plausible analytical situations. Although these tasks provided a consistent study setting, they did not reproduce the domain knowledge, sustained use, collaboration, or consequences involved in real-world decision-making. Participants may behave differently when they have personal stakes in the result or remain accountable for decisions made from a visualization. The findings also may not extend to specialized analyses, larger datasets, or visual forms beyond the statistical charts emphasized by the tasks.

\paragraph{Systems and interaction context}
The main user study examined interaction with ChatGPT only. All participants used the same study-provided ChatGPT Plus account and began a new chat for each task, which standardized access to the system. We did not record whether account-level memory was enabled. The collected interactions contained no apparent references to or other signs of carryover from earlier sessions; nevertheless, we cannot rule out cross-session influence through account-level memory. Model behavior also changes over time, and the observed failures may differ across versions, interfaces, or providers. Our Gemini and Claude comparison broadens the range of observed model behaviors, but replays only the 60 opening prompts and includes no participant interaction or follow-up. It therefore supports claims about codebook coverage and differences among sampled responses, not about how people would use, evaluate, or adapt to those systems.

\paragraph{Analytical perspective}
The authors selected the designs in~\autoref{tab:tasks} as the most effective way to present the expected analytical patterns. \rev{When assessing participant-generated charts, we also accepted alternative designs that followed established visualization design guidelines; charts did not need to match the selected design exactly to be considered acceptable.} All three coders had experience in visualization design. This background supported judgments about chart validity, usability, and established design conventions, but also oriented the analysis toward concerns recognized within visualization research. Negotiated agreement and consistent application of the finalized codebook reduced coding inconsistencies but did not eliminate this interpretive influence.

\end{revisionblock}

\section{Conclusion}
\begin{revisionblock}
Conversational AI can make visualization generation and interpretation more accessible, but this does not ensure that the resulting analyses are reliable. Through a user study with 20 visualization novices, we identify recurring failures in chart and insight quality, difficulties in specifying requests, limited independent verification, and repair attempts that often fail. Our exploratory role analysis further shows that Delegator and Decision-maker sessions were associated with different failure patterns, while neither role had detectably fewer overall design flaws. Applying the codebook to matched Gemini and Claude responses reveals different system-specific behaviors, showing that consequential failures were not limited to ChatGPT. Together, these findings motivate five design recommendations for making AI-generated visualizations easier to critique, inspect, verify, and repair while preserving users' control over the analysis.
\end{revisionblock}


\bibliographystyle{ACM-Reference-Format}
\bibliography{reference}


\end{document}